# Quantum Optimization for Optimal Power Flow: CVQLS-Augmented Interior Point Method

Farshad Amani, *Graduate Student Member, IEEE*, Amin Kargarian, Senior *Member, IEEE*

***Abstract*— This paper presents a quantum-enhanced optimization approach for solving optimal power flow (OPF) by integrating the interior point method (IPM) with a coherent variational quantum linear solver (CVQLS). The objective is to explore the applicability of quantum computing to power systems optimization and address the associated challenges. A comparative analysis of state-of-the-art quantum linear solvers—Harrow-Hassidim-Lloyd (HHL), variational quantum linear solver (VQLS), and CVQLS—revealed that CVQLS is most suitable for OPF due to its stability with ill-conditioned matrices, such as the Hessian in IPM. To ensure high-quality solutions, prevent suboptimal convergence, and avoid the barren plateau problem, we propose a quantum circuit parameter initialization technique along with a method to guide the IPM along the central path. Moreover, we design an ansatz tailored for OPF, optimizing the expressibility and trainability of the quantum circuit to ensure efficient convergence and robustness in solving quantum OPF. Various optimizers are also tested for quantum circuit parameter optimization to select the best one. We evaluate our approaches on multiple systems to show their effectiveness in providing reliable OPF solutions. Simulations for the 2-bus system are conducted on a commercial IBMQ quantum device, while simulations for the other larger cases are performed using the IBM quantum simulator. While promising, CVQLS is limited by current quantum hardware, especially for larger systems. We use a quantum noise simulator to test scalability.**

*Parameters and Functions:*

| | |
|---|---|
| $B$ | Susceptance matrix |
| $C_{\text{coherent}}^{(k,q)}$ | Global cost function in CVQLS |
| $C_{\text{local}}$ | Local cost function used in CVQLS |
| $C_{H_{K_l}}$ | Controlled unitary applying submatrix $H_{K_l}$ |
| $f_k$ | IPM objective function value |
| $H_k$ | IPM Hessian matrix |
| $J_k$ | IPM Jacobian matrix |
| $k$ | Index for IPM iterations |
| $q$ | Index for CVQLS iterations |
| $r_k$ | Residual vector in IPM |
| $U_{r_k}$ | Unitary that prepares the quantum state $\|-r_k\rangle$ |
| $U_\alpha$ | Unitary encoding the probability vector $\sqrt{\alpha}$ |
| $V(w_k^{(q)})$ | Parameterized variational circuit |
| $Y$ | Bus admittance matrix |
| $\alpha_l$ | Coefficient in the decomposition of $H_k$ as a convex combination |
| $\beta_i$ | Coefficient in Pauli decomposition of a Hermitian matrix |
| $\lambda_k$ | Regularization parameter |
| $\mathcal{L}(x,\lambda,\nu)$ | Lagrangian function of the OPF |
| $\mu$ | Barrier parameter controlling the central path |
| $\mu_{-1}^{(l,l')}(w)$ | Normalization term from Hadamard test without controlled-Z gate |
| $\mu_j^{(l,l')}(w)$ | Expectation from Hadamard test with controlled-Z on qubit $j$ |
| $P_D$, $Q_D$ | Active and reactive power demand |
| $\Phi(x,\mu)$ | Log-barrier augmented objective function |
| $\tau$ | Window size for averaging the IPM objective function values |
| $\text{avg}_k$ | Moving average of objective values |
| $\|\Delta x_k\rangle$ | Quantum state representation of Newton direction |
| $\|\Psi_k^{(q)}\rangle$ | Quantum state resulting from variational circuit |

*Variables:*

| | |
|---|---|
| $\Delta x_k$ | Newton direction |
| $P_G$, $Q_G$ | Active and reactive power generation |
| $V$ | Complex voltage vector |
| $s_i$ | Slack variable for inequality transformation |
| $\theta$ | Voltage angle vector |
| $w_k^{(q)}$ | Variational quantum circuit variables |
| $\lambda$, $\nu$ | Lagrange multipliers for equality and inequality constraints |
| $S_\ell$, $S_u$ | Vectors of lower and upper slack variables |
| $\underline{\nu}_k$, $\overline{\nu}_k$ | Dual variables for lower and upper inequality bounds |

## I. INTRODUCTION

ENERGY management is a fundamental aspect of power systems, overseeing the efficient generation, distribution, and consumption of electricity to ensure system reliability and economics. Central to this management is optimizing the power flow (PF) within the grid, which is addressed through optimal power flow (OPF). OPF focuses on optimizing electricity generation to meet demand while minimizing costs. Classical optimization methods, such as linear and nonlinear programming, have traditionally been used to solve OPF. However, these methods can struggle with

This work was supported by the National Science Foundation under Grants ECCS-1944752 and ECCS-2312086.

The authors are with the Department of Electrical and Computer Engineering, Louisiana State University, Baton Rouge, LA 70803 USA (e-mail: famani1@lsu.edu, kargarian@lsu.edu).

scalability and computational efficiency as the complexity of power systems grows [1]–[3].

Modernizing power systems—driven by new technologies and a shift from monopoly-based structures to restructured markets—has increased the demand for faster and more efficient OPF solutions [4]. While classical OPF methods were sufficient for traditional systems, modern grids require optimization approaches that can scale with growing complexity [5]. Fast OPF is especially important for short-term operations like unit commitment, $n-1$ security, and resilience [6]. Classical algorithms often exhibit cubic or polynomial time even in the best case [7]. Achieving linear or logarithmic scaling would represent a significant advancement, enabling significantly faster OPF solutions.

Several methods are used to solve OPF, each with different time complexities. Classical techniques like Newton-Raphson have a computational time of $O(N^2)$ per iteration due to linear system solves, and despite their quadratic convergence, they may struggle with convergence in complex systems [8]. Linear and quadratic programming methods offer simplicity but typically require $O(N^3)$ time [9]. Machine learning approaches, including deep and reinforcement learning, can achieve solution times of $O(N)$ after training, though the training phase often demands $O(N^3)$ or more [10].

The interior point method (IPM) is recognized for its effectiveness in solving optimization problems with linear or nonlinear constraints. IPM is one of the most commonly used techniques for solving OPF. It operates by iteratively solving linear equation systems using the Newton-Raphson method to progressively approach the optimal solution [11]. This iterative process contributes to a computational complexity of approximately $O(N^{3.5})$ [9]. Although IPM is effective for handling complex and large-scale power systems, its performance can be hindered by the repeated need to solve these computationally intensive linear systems, which often become ill-conditioned or nearly singular, making them challenging to address using classical computing resources.

Emerging quantum computing techniques offer the potential for exponential speedup in OPF by leveraging advanced quantum solvers and optimization algorithms. Despite current challenges with noise and errors, quantum computing provides a promising path to improve OPF computational efficiency. A major portion of OPF runtime is spent repeatedly solving large linear systems. Quantum linear solvers can accelerate this process and enhance the scalability of power system optimization. Notable approaches include the Harrow-Hassidim-Lloyd (HHL) algorithm [12], variational quantum linear solver (VQLS) [13], and coherent VQLS (CVQLS) [14].

HHL provides exponential speedup over classical methods but requires fault-tolerant quantum hardware, making it impractical for current noise-sensitive devices [15]. VQLS, by contrast, is suited for noisy intermediate-scale quantum (NISQ) devices, combining quantum and classical processing through a variational approach [16]. Its performance depends heavily on the choice of ansatz, which affects both convergence and accuracy [17]. Research into adaptive ansatz design and machine learning–guided selection has aimed to improve this [18]. CVQLS builds on VQLS by preserving coherence and reducing quantum-classical interactions, enhancing scalability and noise tolerance [14]. However, fully leveraging CVQLS still depends on advances in quantum hardware. Each quantum solver offers trade-offs suited to different problem types and device capabilities.

Applying these quantum methods as linear solvers has the potential to revolutionize multiple fields, including artificial intelligence and optimization, which are fundamental across numerous disciplines. While VQLS has found applications in areas such as machine learning [19], finite element analysis [20], and PF analysis [21], its use in solving OPF poses unique challenges. These challenges include handling sparse, ill-conditioned matrices that can require more computational resources and potentially reduce solution accuracy.

This paper addresses these challenges by exploring the application of quantum computing to OPF, focusing on the practical implementation of quantum linear solvers in power system optimization. We provide a detailed analysis of various quantum linear solvers to identify the most suitable one for our application. We formulate and solve a quantum OPF (Q-OPF) problem using a quantum IPM (QIPM), incorporating a CVQLS at its core. We propose several techniques to address the challenges of QIPM and improve its performance in solving Q-OPF. These include developing a tailored Q-OPF ansatz, optimizing ansatz parameter initialization to impact the search space effectively, and modifying the IPM central path to ensure achieving high-quality solutions even in the presence of quantum computing noise and errors. Additionally, we discuss the performance of available quantum hardware and linear solver packages under real-world conditions and highlight the challenges associated with their implementation.

## II. Quantum Linear Solvers

Solving OPF using IPM relies on solving systems of linear equations. In this context, we explore the practicality and applicability of three well-known quantum linear solvers—HHL, VQLS, and CVQLS—that can be integrated into IPM.

### *A. HHL*

HHL is one of the seminal algorithms for solving systems of linear equations on quantum devices. It is recognized for its minimal dependency on system size and operation on a logarithmic time scale. However, the depth of the quantum circuit in HHL scales with both the size of the linear system and the condition number of matrix $A$ (in $Ax = b$). The overall circuit depth comes from quantum phase estimation, controlled rotations, and inverse quantum Fourier transforms [22].

A significant drawback of HHL for real-world applications is its efficiency, which heavily depends on the system's condition number. As the condition number increases, the algorithm's running time increases, making HHL potentially slower for larger matrices. Additionally, HHL is highly sensitive to noise and errors, limiting its practicality on NISQ devices. Moreover, decoding the solution from a quantum state to classical digits requires quantum tomography, which introduces a linear dependency on system size and, therefore, eliminates the logarithmic advantage of HHL.

### *B. VQLS*

VQLS is a recently developed hybrid algorithm that addresses many HHL limitations, albeit with a dependency on classical computers [13]. VQLS is more resilient to noise and errors, making it suitable for NISQ computers [17]. While the condition number influences the quantum circuit depth, the overall number of quantum gates generally increases with the complexity of the matrix and its condition. The number of cost function evaluations needed to optimize parameters in the variational ansatz also increases with system size. The complexity of these evaluations is often tied to the properties of matrix $A$ and the specific optimization algorithm used. Regarding VQLS convergence, larger, ill-conditioned systems may require more iterations to reach a solution. Sometimes, the algorithm could diverge or get stuck in a barren plateau, where progress toward the solution significantly slows or stops.

### *C. CVQLS*

A variant of VQLS, known as CVQLS, addresses some VQLS limitations with a tradeoff of increased classical computation to guide the quantum circuit. CVQLS reduces the overhead of quantum-classical interaction by relying more on classical computation. The performance and convergence of VQLS are strongly influenced by the choice of the variational ansatz, which defines the structure of the quantum circuit. A poor selection of ansatz can result in slow convergence or suboptimal solutions. In contrast, CVQLS introduces coherence-preserving techniques that make it more resilient to quantum noise, rely less on repeated quantum measurements, have a lower circuit depth, and generally exhibit better convergence with a lower risk of encountering barren plateaus [23]. These characteristics enhance CVQLS scalability and effectiveness, especially for handling ill-conditioned and large systems. By reducing the quantum-classical communication overhead, CVQLS improves efficiency regarding quantum resource utilization. Furthermore, its coherence-preserving nature leads to more accurate results with fewer quantum resources, making it a powerful tool for future quantum-augmented optimization.

### *D. Comparison of Quantum Linear Solvers for Q-OPF Application*

Table I provides a detailed comparison of HHL, VQLS, and CVQLS [12], [13], [24]. This comparison allows one to assess the strengths and weaknesses of each solver, enabling the selection of the most suitable method based on the specific requirements of the problem at hand. In the subsequent sections, we have compared the performance of VQLS and CVQLS. Based on our analysis of the advantages and disadvantages of these three quantum linear solvers, we have selected CVQLS for the Q-OPF application.

IPM in OPF often encounters ill-conditioned systems of linear equations, particularly as the algorithm progresses over multiple iterations. This worsening condition increases numerical instability, making it more difficult to solve the linear system accurately. CVQLS addresses this challenge by leveraging quantum coherence to mitigate the effects of ill-conditioning. Its ability to handle such systems more robustly makes CVQLS an ideal choice for the Q-OPF application, offering improved stability and performance compared to other quantum solvers. CVQLS offers several key advantages that justify its selection. As demonstrated in Table I, CVQLS exhibits superior scalability compared to HHL and VQLS, making it better suited for larger and more complex systems. Additionally, CVQLS reduces the level of interaction required during the solving process, offering improved performance in noisy environments where interaction with qubits can degrade results. Finally, while both VQLS and CVQLS involve hybrid quantum-classical methods with similar computational times, CVQLS generally requires fewer iterations ($T$), making it more efficient in terms of execution time.

TABLE I:
QUANTUM LINEAR SOLVERS FOR NISQ DEVICES

| | **HHL** | **VQLS** | **CVQLS** |
|---|---|---|---|
| Approach | Direct | Hybrid | Hybrid |
| # Qubits | $O(\log_2 N)$ | $O(\log_2 N)$ | $O(\log_2 N)$ |
| Circuit depth | $O(\text{poly}(\log_2 N, \kappa))$ | $O(\text{poly}(\log_2 N, \kappa))$ | $O(\text{poly}(\log_2 N, \kappa))$ |
| Computational time | $O(\text{poly}(\log_2 N, \kappa))$ | $O(T \times \text{poly}(\log_2 N, \kappa))$ | $O(T \times \text{poly}(\log_2 N, \kappa))$ |
| Scalability | - | + | ++ |
| Noise sensitivity | -- | ++ | ++ |
| Interaction | None | High | Reduced |
| Applicability | - | + | ++ |

$T$ is the number of iterations, and $T_{CVQLS} \leq T_{VQLS}$.

## III. QUANTUM OPF

OPF is a critical problem in power systems engineering aimed at optimizing the operation of electrical power networks. The objective is to minimize the generation cost while satisfying various physical and operational constraints, including power balance, voltage limits, and generator limits. IPM is one of the most popular and effective methods for solving OPF, iteratively refining the solution by solving a series of linear equations [25]. We integrate CVQLS into IPM to solve the linear system of equations encountered at each iteration to determine the Newton direction.

### *A. OPF Formulation*

We have used the standard ACOPF formulation, as in MATPOWER [26], which is a nonlinear and nonconvex optimization problem. The objective function is to minimize the generation cost. The constraint includes the standard AC power flow equations, branch flow limits, bus voltage magnitude and angle limits, and generator capacity limits. We have also used the standard DCOPF, as in MATPOWER [26], which is a convex approximation of ACOPF considering only active power and voltage angles. The constraint includes the standard power balance equations, line flow limits, bus voltage angle limits, and generator capacity limits.

**(1)**

$$
\begin{aligned}
&\min_{P_G, Q_G, V, \theta} \quad f(P_G) \\
&\text{s.t.} \\
&P_G - P_D = \text{Re}\left\{V \cdot (YV)^*\right\}, \\
&Q_G - Q_D = \text{Im}\left\{V \cdot (YV)^*\right\}, \\
&V_{\min} \leq |V| \leq V_{\max}, \\
&P_{G,\min} \leq P_G \leq P_{G,\max}, \\
&Q_{G,\min} \leq Q_G \leq Q_{G,\max}, \\
&\theta_{\min} \leq \theta \leq \theta_{\max}, \\
&\theta_{ref} = 0
\end{aligned}
$$

**(2)**

$$
\begin{aligned}
&\min_{P_G, \theta} \quad f(P_G) \\
&\text{s.t.} \\
&P_G - P_D = B\theta, \\
&P_{G,\min} \leq P_G \leq P_{G,\max}, \\
&\theta_{\min} \leq \theta \leq \theta_{\max}, \\
&\theta_{ref} = 0
\end{aligned}
$$

### *B. Interior Point Method*

IPM solves OPF by iteratively finding the Newton direction through a series of transformations and optimizations [27]. We briefly overview the steps involved in IPM and then explain how they can be adapted to a QIPM.

*Step 1: Converting Inequalities into Equalities*: Inequality constraints are reformulated as equality constraints by introducing non-negative slack variables $s_l$ and $s_u$ such that $h(x) - s_l = h_l$ and $h(x) + s_u = h_u$, where $s_l, s_u \geq 0$.

*Step 2: Non-negativity as a Logarithmic Barrier:* Non-negativity constraints on the slack variables are penalized in the objective function using a logarithmic barrier function.

$$
\Phi(x, \mu) = f(P_G, Q_G) - \mu(\ln(S_l) + \ln(S_u)) \tag{3}
$$

*Step 3: Forming an Unconstrained Optimization Function:* An unconstrained optimization problem is formulated as a Lagrangian function, which incorporates both modified objective function and equality constraints:

$$
\begin{aligned}
\mathcal{L}(x, \lambda, \nu) &= \Phi(x, \mu) - \sum_j \lambda_j (g_j(x) - b_j) \\
&- \sum_k \overline{\nu}_k(-h_k(x) + \overline{h}_k - S_u) - \sum_k \underline{\nu}_k(h_k(x) - \underline{h}_k - S_l)
\end{aligned} \tag{4}
$$

*Step 4: Newton Direction Matrix:* The Newton direction $\Delta x$ is obtained by solving perturbed Karush-Kuhn-Tucker (KKT) conditions, which require solving a linear system of equations. To reduce the size of the Newton matrix, we exploit the sparsity of the Hessian $H_k$ and the Jacobian matrices and use matrix factorization techniques. The reduced KKT system is:

$$
\begin{bmatrix} H_k & J_k^T \\ J_k & 0 \end{bmatrix} \begin{bmatrix} \Delta x \\ \Delta \lambda \end{bmatrix} = - \begin{bmatrix} r_k \\ c_k \end{bmatrix}, \tag{5}
$$

To reduce the full perturbed KKT system to the form in (5), we analytically eliminate the slack and dual variables using their respective residual expressions. This yields a smaller Newton system involving only the primal variables and the multipliers associated with the equality constraints. In the reduced system, the following expressions define the Hessian and residual terms [26], [28]:

$$
\begin{aligned}
H_k &= \nabla_x^2 \mathcal{L}_\mu + \mu \nabla h(x)^\top \left(S_\ell^{-2} + S_u^{-2}\right) \nabla h(x), \\
r_k &= \nabla_x \mathcal{L} + \nabla_x h(x)^\top \Big[\mu \left(S_u^{-2} \nabla_{\overline{\nu}_k} \mathcal{L} - S_\ell^{-2} \nabla_{\underline{\nu}_k} \mathcal{L}\right) \\
&\quad + \nabla_{s_\ell} \mathcal{L} - \nabla_{s_u} \mathcal{L}\Big]
\end{aligned} \tag{6}
$$

Once $\Delta x$ is obtained, the updates for the slack variables and their associated dual variables are computed as:

$$
\begin{aligned}
&\Delta s_\ell = \nabla h(x) \Delta x - \nabla_{\underline{\nu}_k} \mathcal{L}, \quad \Delta \underline{\nu}_k = -\mu S_\ell^{-2} \Delta s_\ell - \nabla_{s_\ell} \mathcal{L}, \\
&\Delta s_u = -\nabla h(x) \Delta x - \nabla_{\overline{\nu}_k} \mathcal{L}, \quad \Delta \overline{\nu}_k = -\mu S_u^{-2} \Delta s_u - \nabla_{s_u} \mathcal{L}
\end{aligned} \tag{7}
$$

The smaller reduced system of equations maintains the essential structure for finding the Newton direction while minimizing computational complexity. The next section explains using CVQLS as the quantum linear solver to determine the Newton direction $\Delta x$.

### *C. CVQLS for Newton Direction*

Building on CVQLS advantages, we use it as the quantum linear solver to determine the Newton direction. By leveraging the coherence-preserving capabilities of CVQLS, we can improve the efficiency and accuracy of solving linear systems required to determine the Newton direction.

To solve the linear system $H_k \, \Delta x_k = -r_k$ using CVQLS, it is necessary to represent the Hessian matrix $H_k$ in a quantum state. Several methods exist for encoding classical matrices into quantum states or decomposing them into unitary operators. Each method has distinct characteristics, advantages, and limitations depending on the nature of the matrix and the specific quantum algorithm being applied. Table II provides an overview of some standard encoding methods [29], [30].

TABLE II:
COMPARISON OF QUANTUM MATRIX ENCODING METHODS

| | # Qubits | Complexity | Matrix Type | Limitation |
|---|---|---|---|---|
| Basic encoding | ++ | -- | Arbitrary | High # qubits |
| Amplitude encoding | - | + | Normalized | Arbitrary data |
| Qubitization | + | +++ | Sparse | High complexity |
| Matrix exponentiation | + | ++ | Hermitian | Resource intensive |
| Pauli decomposition | + | + | Hermitian | Scales with matrix size |

Among these encoding methods, we chose Pauli decomposition to encode the classical matrix. Pauli matrices form a complete basis for Hermitian matrices, making them particularly suitable for quantum algorithms that operate on Hermitian operators, like CVQLS. The decomposition of a Hermitian matrix into a linear combination of Pauli matrices allows the representation of the matrix in a form that is naturally compatible with quantum circuits. Furthermore, the scalability of Pauli decomposition through tensor products of Pauli matrices enables the efficient representation of larger matrices, such as those encountered in OPF. This approach simplifies the translation of classical matrix operations into quantum gates, making it an ideal choice for Q-OPF.

We decompose the Hermitian matrix ($H_k^\dagger = H_k$) into a combination of four Pauli matrices as follows:

$$I = \begin{pmatrix} 1 & 0 \\ 0 & 1 \end{pmatrix}, \quad X = \begin{pmatrix} 0 & 1 \\ 1 & 0 \end{pmatrix}, \quad Y = \begin{pmatrix} 0 & -i \\ i & 0 \end{pmatrix}, \quad Z = \begin{pmatrix} 1 & 0 \\ 0 & -1 \end{pmatrix} \tag{8}$$

Assume we have a random Hermitian matrix $H_k = \begin{pmatrix} H_k^{11} & H_k^{12} \\ H_k^{21} & H_k^{22} \end{pmatrix}$. We form a linear combination of Pauli matrices as follows [30]:

$$\beta_0 I + \beta_1 X + \beta_2 Z + \beta_3 Y = \begin{pmatrix} \beta_0 + \beta_2 & \beta_1 - i\beta_3 \\ \beta_1 + i\beta_3 & \beta_0 - \beta_2 \end{pmatrix} = H_k \tag{9}$$

Given a matrix $H_k$, coefficients $\beta$ can be determined. For larger matrices, the combination of tensor products of Pauli matrices can be used to represent $H_k$. For a 4×4 matrix, the combination are:

$$\begin{aligned} &\beta_0'(I \otimes I) + \beta_1'(I \otimes X) + \beta_2'(I \otimes Y) + \beta_3'(I \otimes Z) \\ &\quad + \beta_4'(X \otimes I) + \ldots + \beta_{15}'(Z \otimes Z) \end{aligned} \tag{10}$$

where $\otimes$ denotes the tensor product, and the coefficients $\beta_i'$ correspond to the contribution of each Pauli matrix combination in the representation of $H_k$.

Consider a matrix $H_k$ of size $2^n \times 2^n$, that is expressed as a linear combination of $L$ unitary matrices $H_{k_0}, H_{k_1}, \ldots, H_{k_{L-1}}$. Each of these unitary matrices, $H_{k_l}$, must be efficiently implemented through a quantum circuit that acts on $n$ qubits. The coefficients $\alpha_l$ in this linear combination, represented in equation (9), are part of a normalized and positive probability distribution. The complex phase of each coefficient $\alpha_l$ can be absorbed into the corresponding unitary matrix $H_{k_l}$, leaving us with a vector of positive values for $\alpha_l$.

$$\alpha_l \geq 0 \quad \forall l, \quad \sum_{l=0}^{L-1} \alpha_l = 1 \tag{11}$$

where $L = 2^m$ and $m$ is a positive integer. This normalization ensures that $\alpha_l$ forms a probability distribution, as the linear problem is defined up to a scaling factor.

Next, we introduce a unitary circuit $U_\alpha$ that embeds the square root of $\alpha$ into the quantum state $|\sqrt{\alpha}\rangle$ on $m$ ancillary qubits as:

$$|\sqrt{\alpha}\rangle = U_\alpha|0\rangle = \sum_{l=0}^{L-1} \sqrt{\alpha_l}|l\rangle \tag{12}$$

Where $|l\rangle$ denotes the computational basis state of the ancillary qubits.

For each unitary matrix $H_{k_l}$, we define a controlled unitary operation $C_{H_{k_l}}$ that acts on both the system and the ancillary qubits. The controlled unitary applies $H_{k_l}$ only when the ancillary qubits are in the corresponding basis state $|l\rangle$. This operation can be described as:

$$C_{H_{k_l}}|j\rangle|l'\rangle = \begin{cases} (H_{k_l} \otimes I)|j\rangle|l\rangle & \text{for } l' = l \\ |j\rangle|l'\rangle & \text{for } l' \neq l \end{cases} \tag{13}$$

In this case, the unitary $H_{k_l}$ is applied to the system qubits when the ancillary qubits are in the state $|l\rangle$, while the state remains unchanged if the ancillary qubits are in any other basis state. This construction allows for the selective application of the unitaries in the linear combination based on the state of the ancillary system.

CVQLS approximates the solution $|\Delta x_k\rangle$ with a variational quantum circuit $V(\omega)$:

$$|\Delta x_k\rangle = V(w)|0\rangle \tag{14}$$

The given complex vector $-r_k$ must be generated by a unitary operation $U_{r_k}$ applied to the ground state of $n$ qubits to have the quantum state $|r_k\rangle$:

$$|-r_k\rangle = U_{r_k}|0\rangle \tag{15}$$

To solve $H_k \Delta x_k = -r_k$ we prepare a quantum state $|\Delta x_k\rangle$ such that $H_k|\Delta x_k\rangle$ is proportional to $|-r_k\rangle$, i.e.,

$$|\Psi\rangle := \frac{H_k|\Delta x_k\rangle}{\sqrt{\langle \Delta x_k|H_k^\dagger H_k|\Delta x_k\rangle}} \approx |-r_k\rangle \tag{16}$$

where $\omega$ are the variational parameters. The cost function to be minimized is:

$$C_{\text{coherent}}^{(k,q)} = 1 - |\langle -r_k|\Psi_k^{(q)}\rangle|^2 + \lambda_k \mathcal{C}(V(w_k^{(q)})) \tag{17}$$

$\mathcal{C}(V(\omega_k))$ denotes the coherence measure of the variational circuit at CVQLS iteration $q$ within the $k^{th}$ IPM iteration, and $\lambda_k$ is the regularization parameter. Minimizing $C_{coherent}$ with respect to $\omega$ yields the optimal parameters for the variational circuit.

The global cost function in (17) captures the full overlap fidelity between the normalized state $|-\Psi_k\rangle$ and the target $|-\mathrm{r}_k\rangle$. However, it requires swap tests or inner-product estimations, which can be challenging on near-term quantum hardware. To address this, a local variant of the CVQLS cost function can be constructed by decomposing the matrix $H_k = \sum_l \alpha_l \; l \, H_{k_l}$ and approximating the overlap $\langle -r_k \, |H_k \mid \Delta x_k\rangle$ through a series of local Hadamard tests between $H_{k_l}|x\rangle$ and $H_{k_{l'}}$ $|b\rangle$ on individual qubits. This results in a simplified cost function of the form:

$$C_{\text{local}}(w) = \frac{1}{2} - \frac{1}{2n} \cdot \frac{\sum_{l,l'} \alpha_l \alpha_{l'} \sum_{j=0}^{n-1} \mathrm{Re}[\mu_j^{(l,l')}(w)]}{\sum_{l,l'} \alpha_l \alpha_{l'} \mathrm{Re}[\mu_{-1}^{(l,l')}(w)]} \tag{18}$$

where $\mu_j^{(l,l')}(w)$ denotes the expectation value obtained from the local Hadamard test with a controlled-Z gate on qubit $j$, and $\mu_{-1}^{(l,l')}(w)$ corresponds to the normalization term (i.e., no control-Z gate). This formulation avoids global state overlaps and enables more efficient estimation on current quantum devices while converging to the same solution [13].

The pseudo-code for solving OPF using CVQLS is shown in Algorithm I. This integrated approach combines the robustness of IPM with the computational power of quantum algorithms through CVQLS, where quantum circuits solve the linear systems and classical computations handle the parameter tuning and overall optimization process. The process is illustrated in Fig. 1, where all classical computations are shown on the right side within a dashed box, while the left side demonstrates the Newton step solution using CVQLS. This is a hybrid approach where the optimization process involves quantum circuit parameter tuning managed by a classical computer. Each IPM iteration needs corresponding iterations of CVQLS to solve the linear equations.

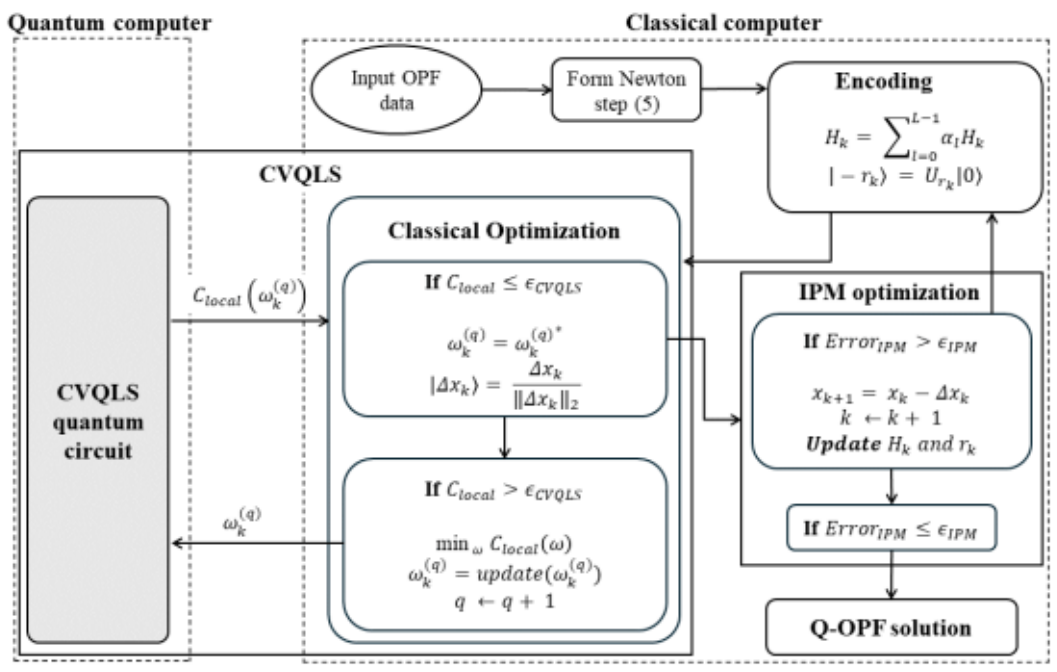

**Fig. 1:** The hybrid approach of solving OPF using CVQLS.

**Algorithm I:** Q-OPF Using CVQLS

1: **Input**: OPF data
2: Form unconstrained OPF optimization (4)
3: Form a system of linear equations (5) to find $H_k \Delta x_k = -r_k$
4: **While** $\epsilon_{IPM} < \epsilon_0$ **or** $k < k^{max}$
  5: Initialize $x_k$ to find classical values
  6: Encode $H_k$ and $-r_k$ into quantum state
  7: Prepare all qubits in the ground state
  8: Apply $U_\alpha$ to ancillary qubits
  9: Apply variational circuit $V(w_k^{(q)})$ to generate $|x(w_k^{(q)})\rangle$
  10: For all $l, l'$, apply controlled unitaries $C_{H_{k_l}}$ and $C_{H_{k_{l'}}}$
  11: Apply Hadamard gates and controlled-Z operators to perform Hadamard tests
  12: Measure ancillary qubit to evaluate $\mu_j^{(l,l')}(w)$ and $\mu_{-1}^{(l,l')}(w)$
  13: Form local cost function $C_{\text{local}}$ as in (18)
    14: **While** $C_{\text{local}} > \epsilon_{CVQLS}$ **and** $q < q^{max}$
      15: Optimize $w_k^{(q)}$ to minimize $C_{\text{local}}$, and update $\Delta x_k$
    16: **End** inner loop
  17: Update $x_{k+1} = x_k - \Delta x_k$
  18: Update $k \leftarrow k + 1$
  19: Calculate $\epsilon_{IPM}$
20: **End** outer loop
**Return**: Q-OPF solution

## IV. Proposed Q-OPF Improvement Techniques

The current quantum linear solvers are primarily designed for proof-of-concept purposes and work under controlled conditions. Significant adjustments are needed across applications, as the problem's physics affects algorithm performance. In CVQLS, challenges include designing effective ansatz structures, handling quantum noise, and achieving stable convergence. Since we use a variational approach, we first outline the key components and optimization strategies tailored to our OPF application. To accelerate CVQLS within the IPM framework, we implement multiple strategies, including sequential parameter initialization, empirical ansatz selection, optimizer choice, and early stopping. Also, to mitigate the difficulties that ill-conditioned linear systems pose for CVQLS, we combine measures including 1) supplying Hermitian KKT matrices to CVQLS after dimension reduction to improve their condition numbers, 2) re-using the optimized parameters from the IPM preceding step to start the variational search near the true solution even when the cost landscape is flat or irregular, and 3) using adaptive optimizers (Adam) and early-stopping thresholds to curb overfitting to noise when exact convergence is unattainable. Together, these strategies lower circuit depth, reduce iterations and runtime, and improve robustness and convergence for ill-conditioned systems.

### A. Varaitional Quantum Optimization Landscape

In variation quantum algorithms, the quantum circuit parameter optimization landscape is influenced by the choice of the variational ansatz and its parameter initialization. These factors determine whether the optimization process encounters barren plateaus or converges to a solution.

Fig. 2 illustrates different optimization landscapes encountered in CVQLS [31]. In Fig. 2(a), no barren plateau is observed, as multiple viable paths lead to the solution. This occurs when the ansatz is well-structured, and parameter initialization is near the optimal region, facilitating smooth convergence. Fig. 2(b) shows an optimization landscape approaching a barren plateau, where several local minima exist, and only a few paths guide the optimization towards the global minimum. Such a landscape can arise when the ansatz is moderately complex or when parameter initialization is less favorable, limiting the number of successful optimization routes. Fig. 2(c) shows a nearly barren plateau with few paths to the solution. The flat landscape indicates an overly complex or poorly chosen ansatz, resulting in limited convergence regions. This often occurs in CVQLS due to inefficient optimization, such as over-parameterization or poor initialization.

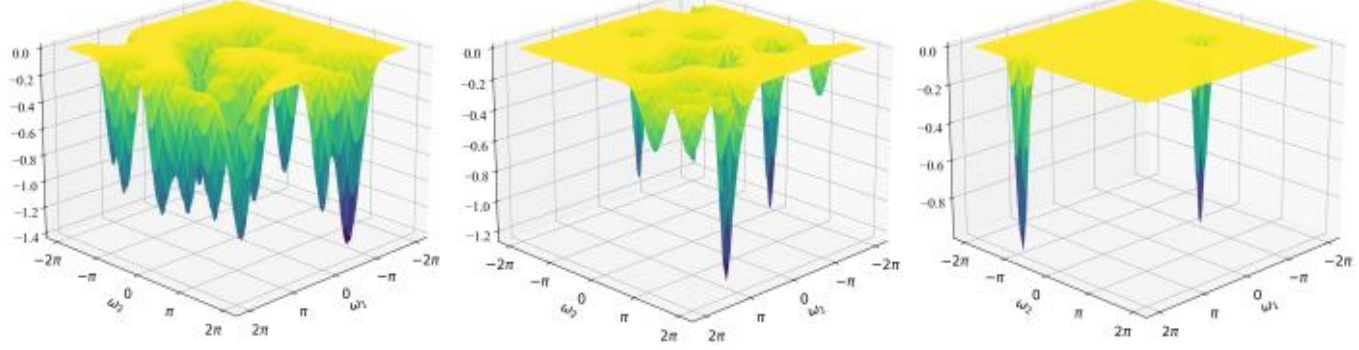

(a) Multiple solution paths (b) Partially barren plateau (c) Barren plateau

**Fig. 2:** Varaitional quantum optimization landscape with different settings.

### B. OPF-Tailored Ansatz

The choice of ansatz is crucial for CVQLS performance. The ansatz is shallow or deep, presenting distinct characteristics and trade-offs. The depth variable controls the number of layers in the ansatz, where each layer typically consists of Ry gates (rotation around the Y-axis) followed by CZ gates (controlled-Z gate). The Ry gate rotates the qubit's state around the Y-axis on the Bloch sphere. It takes a parameter $\phi$ and transforms the qubit state based on that angle. In the ansatz, Ry gates allow continuous tuning of the qubit states to optimize the solution. CZ gate is a two-qubit gate that applies a Z-phase flip (changes the phase of the qubit state) on the second qubit if the first qubit is in the $|1\rangle$ state. CZ gates create entanglement between qubits in the ansatz, enabling complex interactions to solve linear systems.

Depending on the problem, selecting an appropriate ansatz is important for CVQLS [16]. A shallow ansatz reduces the expressivity of the quantum circuit, making it less capable of representing complex functions. Consequently, to accurately approximate the solution of the linear system, a deeper ansatz with higher expressivity might be necessary. However, increasing the depth of the ansatz can introduce trainability challenges. A more complex circuit complicates the optimization landscape, increasing the likelihood of encountering barren plateaus where the cost function gradient is nearly zero. Furthermore, while a deeper ansatz can lead to overfitting, a shallow ansatz might fail to find the solution, resulting in poor convergence [16], [32]. The comparison of a shallow and deep ansatz is given in Table III.

TABLE III:
ANSATZ EFFECTS

| | Shallow | Deep |
|---|---|---|
| Expressivity | Limited | High |
| Trainability | Easy | Hard |
| Overfitting | Low risk | High risk |
| Convergence | Fast | Slow |
| Optimization | Fast | Slow |
| Noise | Less prone | Susceptible to noise |

To identify a suitable ansatz for our application, we conducted a series of evaluations using real linear systems extracted from IPM iterations while solving the OPF problem. We compared multiple ansatz structures and selected the one that consistently produced lower solution errors across various representative scenarios. This empirical approach allowed us to tailor the ansatz to the structure of the problem at hand rather than relying solely on theoretical guarantees.

Ansatz design remains an open and evolving area of research in variational quantum algorithms. Several strategies—from hardware-efficient designs to problem-inspired constructions—have been proposed, each with its trade-offs [16]. Our Q-OPF approach is compatible with future improvements, and more specialized ansatz configurations can be readily integrated as the field progresses.

Two used ansatzes are shown in Fig. 3. Based on our observations, the shallow ansatz Fig. 3(a) shows stable and reliable convergence in our study. However, the deeper ansatz shown in Fig. 3(b) either converges very slowly or fails to effectively reduce the CVQLS cost function. Thus, we suggest the shallow ansatz for the Q-OPF application.

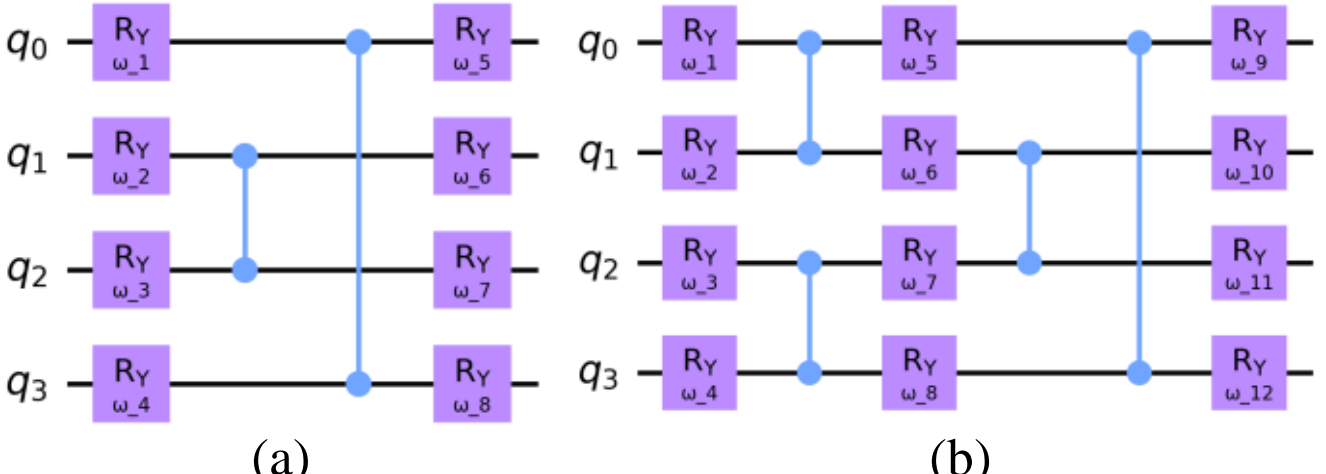

**Fig. 3:** a) Shallow, and b) deep ansatzs.

### *C. CVQLS Parameter Initialization*

roper initialization of ansatz parameters is crucial for CVQLS performance, as these initial values significantly influence the optimization landscape the algorithm explores. Poor initialization might place the optimization process in regions of the landscape where the cost function is flat, known as barren plateaus, leading to slow convergence [32]. A well-chosen initialization can position the algorithm closer to a good solution, enhancing convergence speed and reducing the number of iterations between classical and quantum computing devices [23].

Avoiding local minima is another important role that good initialization plays [33]. A more informed initialization can provide a better starting point, making it easier to reach the global minimum. The challenge of barren plateaus, where the cost function gradient is nearly zero over large regions of the parameter space, is a significant concern in variational quantum algorithms like CVQLS. Proper parameter initialization can help mitigate this issue by starting the optimization in a region where the gradients are not flat, making it easier to find a descent direction [23], [32], [33]. We used two different methods for the initialization:

**Fixed Value Initialization:** Initially, we tested the default parameter initialization in the PennyLane CVQLS package, which sets all parameters to one. While simple, this approach led to poor results. Fixed initialization caused high errors and unstable convergence [24], [33]. In many cases, the optimization failed to converge, resulting in CVQLS divergence. These issues likely stem from limited parameter space exploration, with the algorithm often getting trapped in local minima and failing to reach a viable solution path.

**Sequential Initialization Using Parameters from Previous Iterations:** The second approach involved using the parameters obtained from the final iteration of the previous CVQLS run as the initial parameters for the subsequent OPF iteration. Since the structure of the matrices $H_K$ and $H_{K+1}$ are structurally similar, this method leverages the proximity of solutions in parameter space. It enables optimization to begin from a more informed point, reducing the number of iterations required [33], [34]. While fixed-value initialization was the least effective, sequential initialization based on prior knowledge produced significantly better results. We therefore recommend the second method.

### *D. Quantum Circuit Parameter Optimization Methods*

Optimization methods are crucial for minimizing the CVQLS cost function and tuning circuit parameters. These methods are categorized as gradient-based, quasi-Newton, and non-gradient-based optimization approaches [13], [17], [35].

Quasi-Newton methods approximate second-order information (Hessian) to update parameters. These methods can converge faster than basic gradient descent but are computationally more expensive. For CVQLS, the Broyden–Fletcher–Goldfarb–Shanno (BFGS) algorithm works well for small quantum systems but becomes impractical for larger, more complex ones due to its reliance on second-order approximations. Non-gradient-based optimizers, like COBYLA, use linear approximations instead of gradients, making them suitable for quantum systems with noisy or hard-to-compute gradients. While slower to converge than gradient-based methods, COBYLA is valuable in some CVQLS cases due to its noise resilience. Gradient-based optimizers use gradients to update parameters. Adam adjusts learning rates using first and second-moment estimates, making it efficient for large-scale problems. Based on our case study results, we recommend Adam for CVQLS due to its effectiveness in handling large parameter spaces and speeding up convergence. Table IV provides a comparison between the three quantum circuitry optimizers.

TABLE IV:
QUANTUM CIRCUIT PARAMETER OPTIMIZATION METHODS

| | Gradient-based | Quasi-Newton | Non-gradient-based |
|---|---|---|---|
| **Example** | Adam | BFGS | COBYLA |
| **Gradient Requirement** | Yes | Yes (Hessian approx.) | No |
| **Computation Efficiency** | High (adaptive learning) | Medium (second-order info) | Low (linear approx.) |
| **Convergence Speed** | Fast | Medium | Slow |
| **Applicability** | VQLS and CVQLS | Small systems | Noisy systems |

### *E. QIPM Central Path Direction*

Controlling IPM convergence is critical, especially since Newton's direction may be affected by quantum noise. A key parameter in IPM is $\mu$ , which manages the trade-off between optimality and feasibility. It is associated with the barrier term that penalizes proximity to constraint boundaries. Proper tuning of $\mu$ helps keep iterates within the feasible region, improving numerical stability and preventing ill-conditioning near the boundary.

The parameter $\mu$ also guides the iterates along the central path—a trajectory within the feasible region that IPM follows toward optimality. This path helps keep iterates away from constraint boundaries, where numerical issues are more likely. Proper control of $\mu$ ensures that iterates remain close to the central path, reducing the risk of divergence and instability.

Choosing $\mu \approx 0$ emphasizes finding the optimal solution by rapidly tightening the complementary gap, while a larger $\mu$ ensures the feasibility of the solution. Generally, a slower decrease in $\mu$ may result in more iterations but provides greater stability. Conversely, a more aggressive decrease can speed up convergence but risks instability or divergence if the updates are too large [36].

Various strategies have been proposed to update $\mu$ during IPM iterations [37]. In this study, we adopt MATPOWER's original IPM formulation with some modifications, accounting for using an inexact linear solver. Our results show that classical IPM tends to reduce $\mu$ too quickly. More IPM iterations are needed with inexact solvers, requiring careful control of $\mu$ 's reduction rate. If $\mu$ becomes too small early on, the central path tightens prematurely, bringing iterates near the boundary and increasing the risk of convergence to a local or incorrect solution.

To address this, we monitor the IPM objective function value across iterations. We compute the average of IPM objective function values over $\tau$ consecutive iterations using (19).

$$\mathrm{avg}_k = \frac{1}{\tau} \sum_{i=k-\tau}^{k-1} f_i \quad \forall k = \tau+1, \tau+2, \ldots, k^{\max} \tag{19}$$

(19) allows us to smooth changes in IPM objective function and detect any trends in its behavior. To determine whether IPM objective function values are stabilizing, we compute the relative difference between the average value and the next iteration's objective value, as defined by relDif in (20).

$$\mathrm{relDif}_k = \left| \frac{\mathrm{avg}_k - f_{k+1}}{\mathrm{avg}_k} \right| \quad \forall k = \tau+1, \tau+2, \ldots, k^{\max} \tag{20}$$

By monitoring $relDif_k$, we can dynamically control $\mu$ updates based on the behavior of IPM objective function. If $relDif_k$ falls below a threshold, or if the IPM objective function exhibits a sudden deviation beyond $\pm$ 20% $of$ $\mathrm{avg}_i$ over the last few iterations, we fix $\mu$ to ensure stability. Otherwise, $\mu$ is updated dynamically according to classical IPM. This approach ensures that $\mu$ remains stable during periods of large fluctuations while adapting when convergence is steady. Typically, $\mu$ is only fixed once in the early stages of the OPF iteration to let the solver find the correct central path to the optimal solution. This process is further detailed in Algorithm II, and visual representations are in Fig. 4.

### *F. Complexity Analysis*

The computation in the CVQLS-augmented IPM setup is divided between classical outer loops and a quantum step for solving linear systems. In classical IPM, the most expensive part is solving the Newton system, which takes around $O(n^{3.5})$ time in dense cases but can drop to about $O(n^2)$ when we take advantage of OPF's sparsity and block structure [9]. With CVQLS, only this linear solve step changes, and everything else in IPM stays the same.

Suppose the KKT matrix can be written as a sum of sparse Hermitian unitaries and has a reasonable condition number $\kappa$. In that case, CVQLS can solve it with gate complexity that grows polynomially with $\kappa, 1/\varepsilon$ (the solve accuracy), and $\log(n)$, instead of scaling directly with $n$. The classical parts—like Jacobian setup, barrier term updates, and line search—stay around $O(n)$, and IPM still converges in about $O(log(1/\varepsilon))$ iterations [9], [11], [13]. Therefore, the time complexity under ideal conditions can be estimated as:

$$T_{\text{total}} = O\left( \log\left(\frac{1}{\varepsilon}\right) [\mathrm{poly}(\kappa), \mathrm{poly}(\log n) + n] \right) \tag{21}$$

However, this complexity assumes the availability of fault-tolerant quantum hardware. Therefore, the outcomes presented in this work should be interpreted as theoretical upper-bound demonstrations rather than immediate, practical performance gains.

**Algorithm II:** $\mu$ correction for guiding QIPM central path

**Input**: Objective function of QIPM

1: **While** $k < k^{max}$

2: Run QIPM and store the objective function

3: **If** $relDif_k < \epsilon_{conv}$ **or** $f_k < 0.8 \times avg_i$**or**$f_k > 1.2 \times avg_i$

fix the value of $\mu$ to be the same as iteration number $k$

$k \leftarrow k + 1$

4: **Elseif** $relDif_k > \epsilon_{conv}$ **or** $0.8 \times avg_i < f_k < 1.2 \times avg_i$

update $\mu$

$k \longleftarrow k + 1$

5: **End**

**Return**: Value of $\mu$

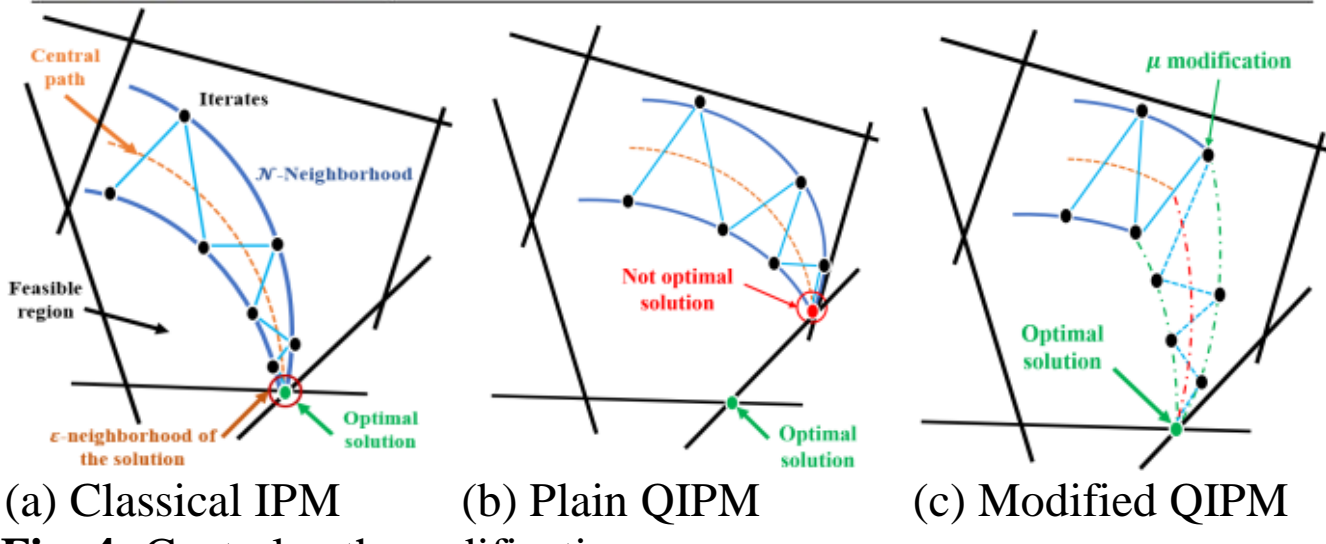

(a) Classical IPM (b) Plain QIPM (c) Modified QIPM

**Fig. 4:** Central path modification.

## V. CASE STUDY

### *A. Simulation Setting*

We tested the proposed approaches on two small systems (3-bus and 5-bus) and two larger systems (118-bus and 300-bus).

CVQLS simulations are conducted using the PennyLane package with Python on a PC with 16 GB of RAM. Adam, BFGS, and COBYLA are used to optimize VQLS and CVQLS quantum circuit parameters. As shown in Table VI, encoding for large systems, such as the 118-bus and 300-bus, encountered out-of-memory errors. This limitation is primarily due to the size of the Hessian matrix, which requires $log_2\, N$ qubits for an $N \times N$ matrix and leads to exponential growth in memory when simulating on classical hardware. In practice, simulating dense Hessians becomes infeasible on standard machines, especially as circuit depth increases with matrix complexity [38]. Therefore, while CVQLS was simulated for the small systems, for the larger systems, we modeled quantum errors and noise using the Qiskit noise simulator and incorporated these into the computations.

### B. OPF vs. PF

Solving linear systems of equations is a key component in both PF and OPF studies. The condition numbers of PF matrices are smaller than those of OPF matrices. Table V presents the condition numbers for PF and OPF problems, highlighting the ill-conditioned nature of OPF matrices compared to those in PF studies. A poorly conditioned matrix complicates the encoding process of matrix $H_k$, leading to higher computational costs and increased susceptibility to noise and errors in calculations. For instance, encoding the first iteration of DCOPF in a 3-bus system needs a circuit with 364 gates:

$$3771.0073 \cdot IIII + 3494.7806 \cdot ZIZI - 3282.7582 \cdot IIZZ + \ldots + (85 \text{ other terms}) + \ldots + 0.3405 \cdot ZXXX - 0.0931 \cdot ZZZX - 0.0139 \cdot ZYYI \tag{22}$$

In contrast, encoding the 3-bus system PF matrix requires three gates.

$$2.6953 \cdot I - 1.3333 \cdot X - 0.2509 \cdot Z \tag{23}$$

The number of controlled Pauli gates for encoding the Hessian matrix ($H_k$) in the last iteration of Newton's step for different systems are displayed in Table VI. In addition, the matrix size in OPF is larger than that of PF, worsening the situation. Table VII shows the sizes of matrices for different systems. These factors make solving OPF with QIPM more complex than solving PF, emphasizing the need for carefully designed and sophisticated approaches to address OPF in a quantum computing setting.

TABLE V:
CONDITION NUMBERS OF PF AND OPF

| | DCPF | ACPF | | DCOPF | | ACOPF | |
|---|---|---|---|---|---|---|---|
| | | min | max | min | max | min | max |
| 3-bus | 3.3 | 3.03 | 4.65 | 1.9e4 | 3e17 | 41 | 1e11 |
| 5-bus | 11.4 | 16.1 | 16.2 | 1.1e4 | 7e19 | 2.2e4 | 3e11 |
| 118-bus | 1.7e4 | 3.1e3 | 3.1e3 | 3e7 | 2e14 | 6.7e3 | 2e9 |
| 300-bus | 1.3e5 | 1.1e5 | 1.1e5 | 4e7 | 5e7 | 1e6 | 3e12 |

TABLE VI:
REQUIRED RESOURCES FOR ENCODING HK

| | 3-bus | 5-bus | 118 bus | 300-bus |
|---|---|---|---|---|
| DCPF | 18 | 62 | Error | Error |
| DCOPF | 364 | 308 | Error | Error |
| ACPF | 3 | 70 | Error | Error |
| ACOPF | 1548 | 1840 | Error | Error |

TABLE VII:
MATRIX SIZE

| | DCPF | ACPF | DCOPF | ACOPF |
|---|---|---|---|---|
| 3-bus | 3 | 2 | 10 | 19 |
| 5-bus | 5 | 5 | 16 | 31 |
| 118-bus | 118 | 181 | 291 | 581 |
| 300-bus | 300 | 530 | 670 | 1339 |

### C. CVQLS vs. VQLS

We have tested Adam and COBYLA optimizers to solve the 16x16 matrix in the 3-bus system during the third iteration of QIPM using VQLS. Although, according to Table IV, Adam is expected to perform the best, as shown in Fig. 7(a), it did not yield satisfactory results and became stuck in a barren plateau situation. COBYLA is applied to the same matrix and demonstrated better results, as illustrated in Fig. 5. However, COBYLA exhibits very slow convergence and the number of quantum-classical iterations varies even when the model is rerun under identical conditions. Fig. 5 illustrates the convergence behavior of VQLS. While the solution approaches the target, fluctuations occur throughout the convergence process. Fig. 5(b) indicates that additional iterations are needed when rerunning the simulation with the same matrix. Also, the solution error is higher in the first case. This suggests that although VQLS performs effectively, achieving consistent results for larger systems can be challenging.

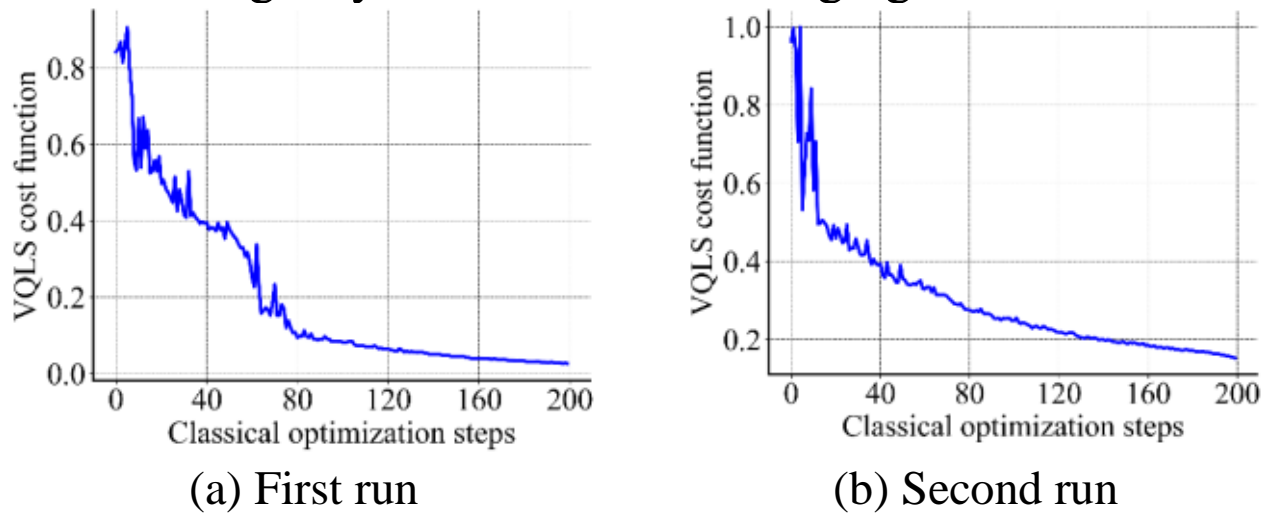

(a) First run (b) Second run

Fig. 5: VQLS cost function of the third QIPM iteration for DCOPF in 3-bus system.

For larger matrices like 32x32 in the 5-bus system during the second QIPM iteration, we had even more challenges. We observed that the computational costs of VQLS to solve the equation $H_k \Delta x_k = -r_k$, increases as the system size grows. However, to find a better behavior, one can change the ansatz and VQLS optimizer and use different ansatz initializations. For example, we have obtained acceptable results for a 32x32 matrix in the second iteration of ACOPF for the 5-bus system using padding to make the matrix Hermitian and the BFGS optimizer. However, this came with trade-offs: the optimization process was time-consuming and exhibited oscillatory behavior in the cost function, as shown in Fig. 6. As shown in Fig. 7(b), we have observed that the COBYLA optimizer encounters a barren plateau situation after a time-consuming process and is unable to find the optimal solution.

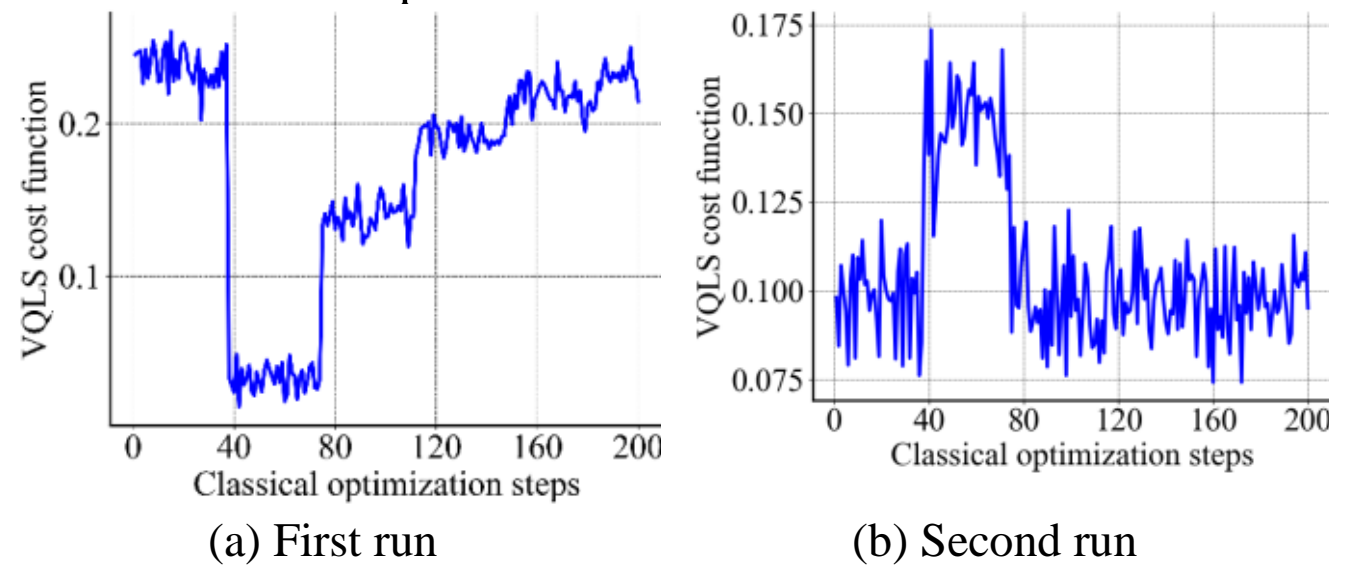

(a) First run (b) Second run

Fig. 6: VQLS cost function of the second QIPM iteration for ACOPF in 5-bus system.

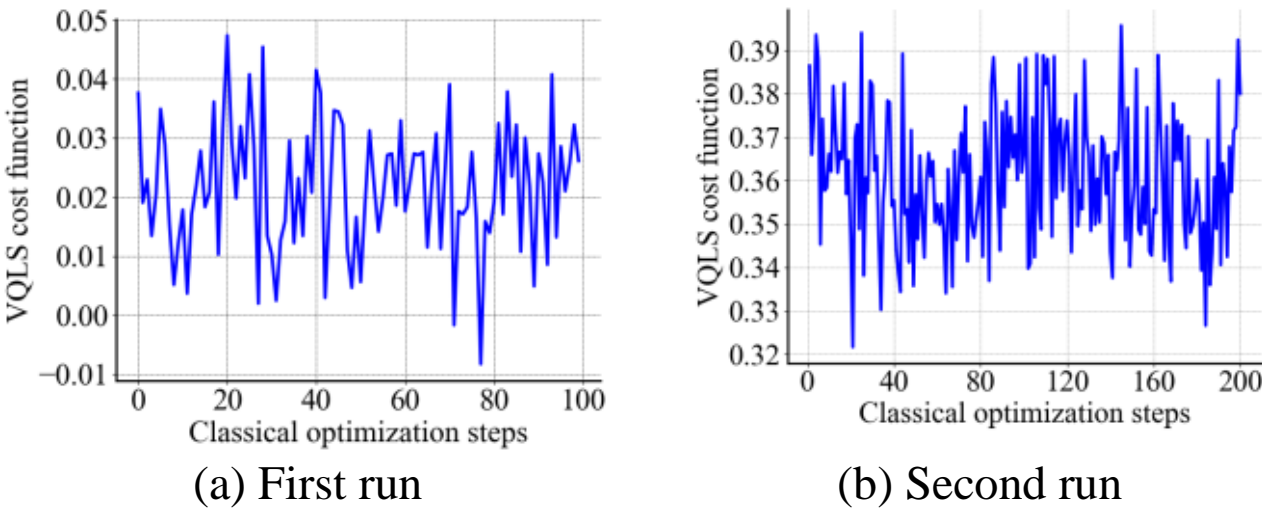

(a) First run (b) Second run

Fig. 7: VQLS cost function (a) in the third iteration of 3-bus system using Adam and (b) in the second iteration of 5-bus system using COBYLA.

These observations highlight the limitations of VQLS in solving OPF matrix equations with high accuracy. The extended computation time and error variability indicate the need for improved quantum algorithms to enhance reliability and efficiency. Such delays present significant challenges for solving ACOPF efficiently. We explored CVQLS to address the challenges faced by VQLS. CVQLS offers better convergence and lower computational overhead by leveraging coherent operations and optimized variational methods. This enhances the feasibility of solving complex problems like OPF, particularly ACOPF. Fig. 8 shows the convergence behavior for 8×8 and 16×16 matrices. For the 32×32 case, CVQLS performance is demonstrated in Fig. 9. The cost function consistently decreases with minor errors, even across repeated runs. These convergence stability and efficiency improvements support the use of CVQLS for Q-OPF. Notably, we exclusively used the Adam optimizer and did not encounter the issues seen with VQLS.

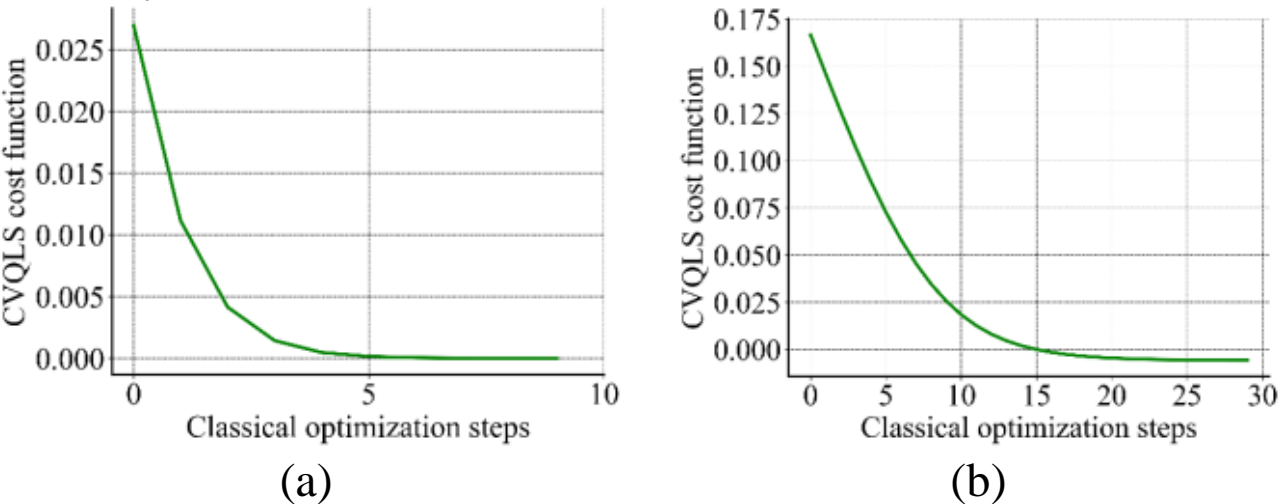

(a) (b)

Fig. 8: CVQLS cost function for (a) ACPF matrix of 5-bus system, and (b) DCOPF of 5-bus system at 5th QIPM iteration.

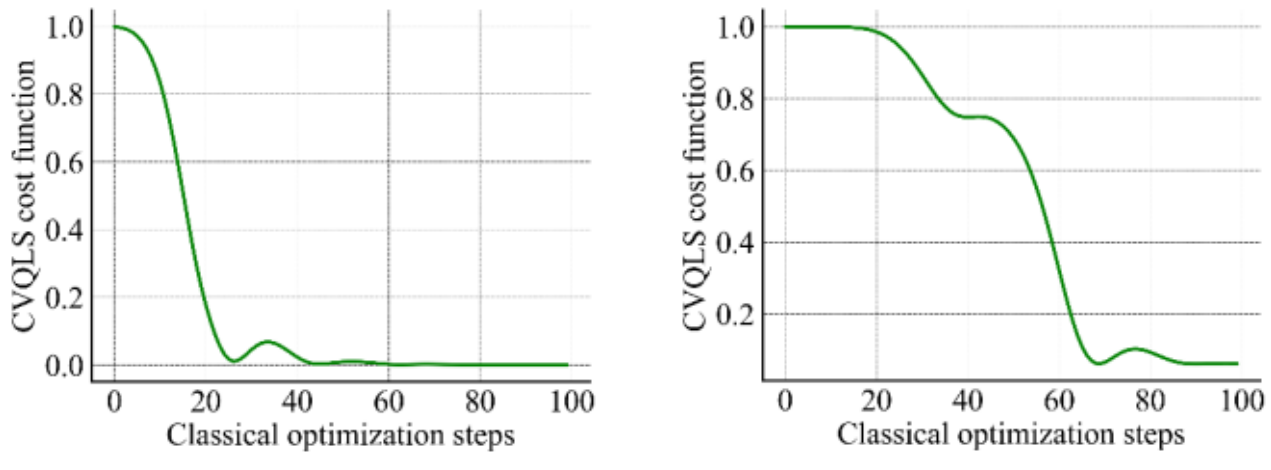

(a) Second QIPM iteration (b) 10th iteration

Fig. 9: CVQLS cost function of the ACOPF in the 5-bus system.

### *D. Q-OPF on Real Quantum Computer*

To demonstrate the application of a real quantum computer and its limitations in solving DCOPF, we used a small 2-bus test system. This system includes a fixed load of 100 MW at Bus 2. The generator is modeled with a quadratic cost function defined as $C(p_g) = 561 + 7.92\, p_g + 0.00156\, p_g^2$, where $p_g$ is constrained in the range of 100 to 400 MW, and the line has a capacity of $\pm 200$ MW. All other system parameters are provided in Fig. 10a.

For this system, the IPM formulation of the DCOPF problem resulted in a linear system of size 7x7, which we padded 8x8 to meet the input size requirement of the CVQLS quantum solver. We used PennyLane's CVQLS module and connected it to IBMQ to run the quantum circuits on real quantum hardware. The circuit shown in Fig. 10b represents the circuit executed on real quantum hardware.

The CVQLS circuit uses four qubits: qubits 0–2 represent the solution vector $x$, and qubit 3 is the flag qubit. The work qubits are initialized using Hadamard and $R_Y$ gates to encode the right-hand side vector $b$. The flag qubit, initialized with a Hadamard gate, controls two variational ansatz blocks $|\Psi\rangle$, which approximate the inverse of matrix $A$. A final Hadamard and measurement on the flag qubit enable postselection. The solution is extracted from the amplitudes of the work qubits, conditioned on measuring the flag in state $|1\rangle$.

The CVQLS method is a hybrid algorithm that requires frequent interaction with a classical optimizer to update variational parameters. As shown in Fig. 10b, each IPM iteration required multiple circuit executions. Since current IBMQ devices lack native hybrid support, each circuit had to be manually submitted, often taking up to three hours per iteration depending on queue length. To reduce delays, we used IBMQ's backend selector to target the least busy device. Note that each CVQLS circuit execution took about 0.8–1 second with 8192 measurement shots.

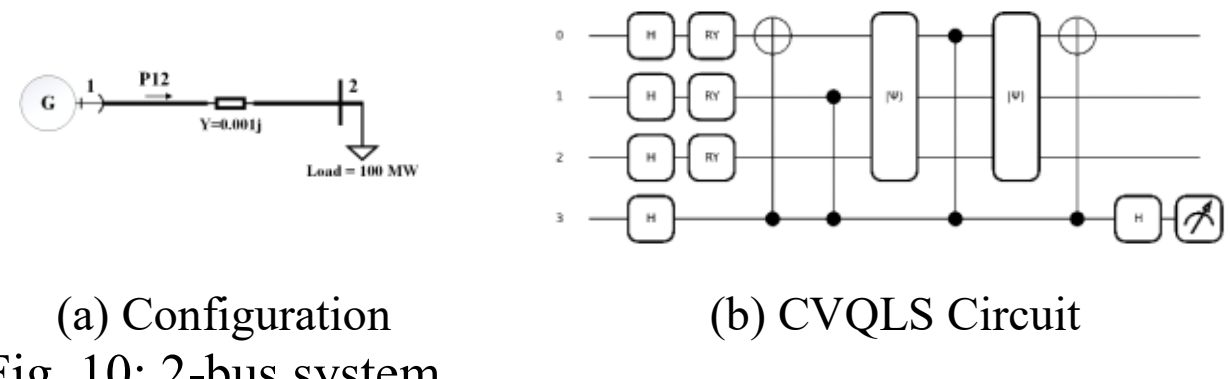

(a) Configuration (b) CVQLS Circuit

Fig. 10: 2-bus system.

Fig. 11a shows QOPF results for the 2-bus DCOPF case. After 12 iterations, the generator output reached 105.3 MW, with a 5% error—reasonable given hardware noise and limitations. Fig. 11b illustrates the convergence trend during iteration 5. We stopped the process manually after 16 iterations, as further improvements became marginal and the solution was already satisfactory.

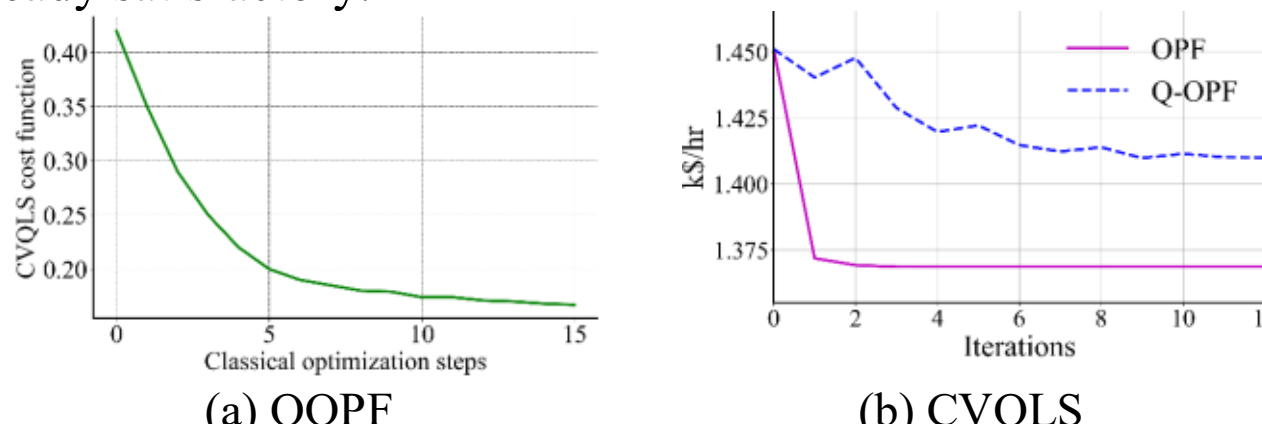

(a) QOPF (b) CVQLS

Fig. 11: Q-OPF objective function of 2-Bus system and CVQLS system.

We executed 192 quantum circuits on real IBMQ hardware for 12 QIPM iterations. This made the overall process highly time-intensive. Moreover, due to significant quantum noise, the accuracy of the results was limited. As a result, real quantum hardware was only used for the 2-bus case, while quantum simulators were used for the remaining test scenarios.

### *E. Q-OPF Simulation with PennyLane*

We ran DCOPF and ACOPF problems using MATPOWER with a classical IPM solver and used the results to compare the Q-OPF outcomes with the proposed QIPM solver. The quantum circuit used for solving the DCOPF of a 3-bus system is shown in Fig. 12. This circuit consists of six qubits: qubits 0-3 serve as work qubits representing the solution vector $x$, qubit four functions as the postselection (control) qubit to enforce coherent conditioning, and qubit five assists with controlled operations and measurements. The circuit is significantly more complex than the one presented in Fig. 10b. Moreover, the depth of the variational block $\psi$, illustrated in (22), is considerably greater, leading to higher implementation error on current quantum hardware. Therefore, we executed this version using a quantum simulator.

In our simulation, CVQLS with the Adam optimizer is used as the quantum linear solver for Q-OPF. Figs. 13 and 14 show that QIPM follows a trajectory similar to classical IPM and performs well on both DCOPF and ACOPF. Despite ACOPF's nonlinear complexity, the approach achieves optimal results, aided by the enhancements in Section IV. More QIPM iterations are needed for the 5-bus system than for the 3-bus case to reach similar objective values.

This suggests that system type and size affect convergence behavior. Increasing QIPM iterations ($k_{IPM}$) can yield satisfactory results, especially for larger systems, which may require more iterations. Although Q-OPF may involve more iterations than classical OPF, our focus is on reducing time per iteration; thus, a higher count does not imply slower convergence. We tested the 3- and 5-bus systems on the PennyLane simulator, but could not evaluate larger cases due to its limited capacity. Additional simulations are needed to assess QIPM's performance on larger and nonlinear ACOPF problems. With ongoing advances in quantum technology, we remain optimistic about extending Q-OPF to larger systems.

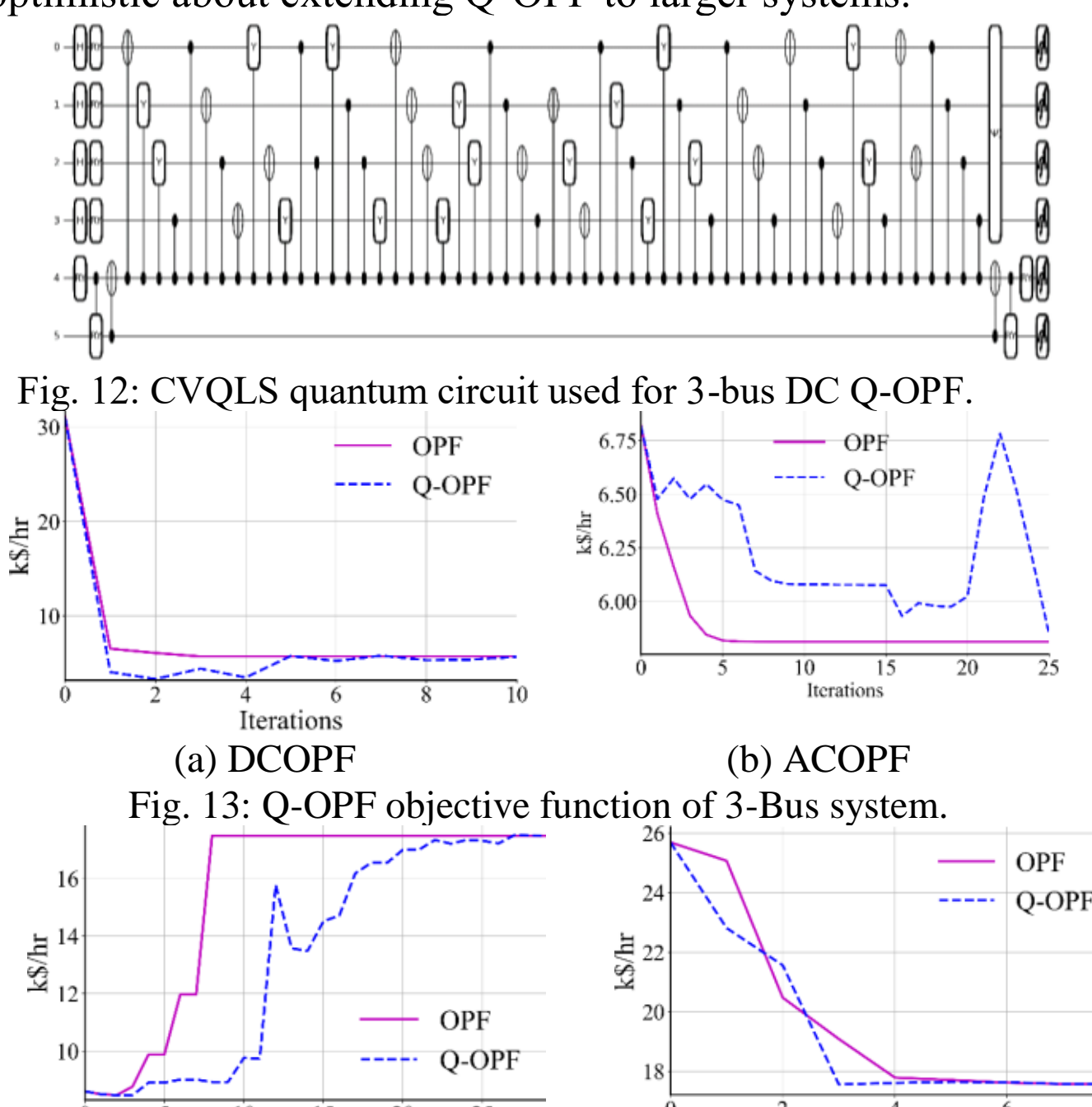

Fig. 12: CVQLS quantum circuit used for 3-bus DC Q-OPF.

(a) DCOPF (b) ACOPF

Fig. 13: Q-OPF objective function of 3-Bus system.

(a) DCOPF (b) ACOPF

Fig. 14: Q-OPF objective function of 5-Bus system.

### *F. Noise Simulation for Large Systems*

To assess performance on larger systems despite hardware limits, we simulate quantum execution and test Q-OPF on IEEE 118- and 300-bus systems. We use the Qiskit noise simulator to model errors typically encountered in real quantum computers. This simulator introduced noise into our linear system's matrix $H_k$ and vector $r_k$. We still use Algorithm II to adjust $\mu$ to ensure QIPM remains stable and avoids divergence.

The DCOPF and ACOPF results for the 118-bus and 300-bus systems are shown in Figs. 15 and 16, respectively. The proposed approach performs effectively, even in the presence of errors and noise. QIPM achieves a solution that closely approximates the reference results obtained by the classical IPM. However, QIPM takes more iterations to converge, which can be attributed to the noise affecting the calculations, thereby preventing the Newton direction from rapidly converging in fewer iterations.

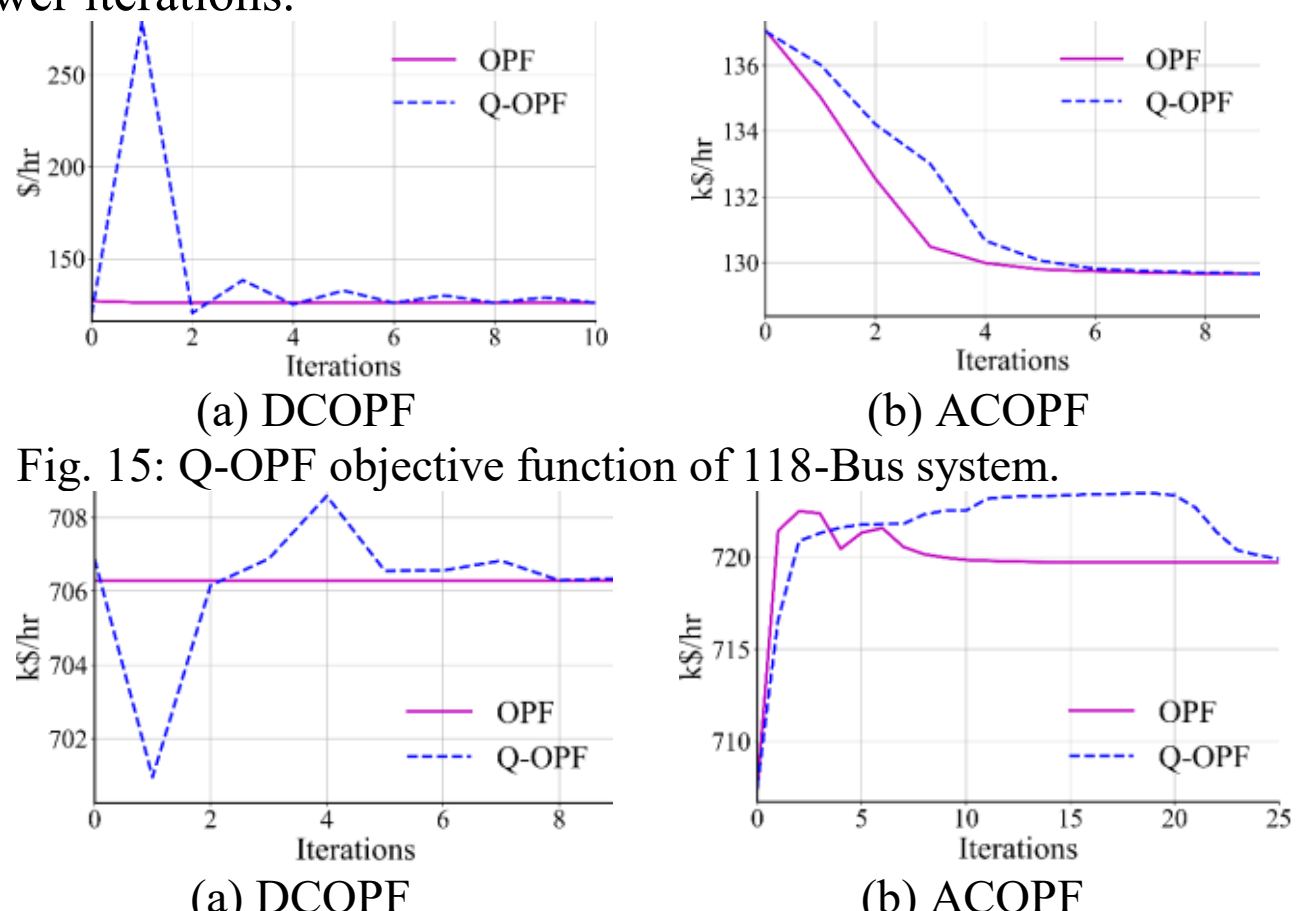

(a) DCOPF (b) ACOPF

Fig. 15: Q-OPF objective function of 118-Bus system.

(a) DCOPF (b) ACOPF

Fig. 16: Q-OPF objective function of 300-Bus system.

### *G. Q-OPF without $\mu$ Correction*

To evaluate the effect of omitting the proposed $\mu$-modification in Q-OPF, we solved 1,000 random load cases in which each bus demand was independently scaled by a uniform factor between 0.8 and 1.2 times its nominal value. Without the $\mu$-correction, a large fraction of runs violated either optimality or feasibility conditions; the exact percentages are reported in Table VIII. In every test system, more than 60% of the simulations failed to converge.

TABLE VIII: Total Failure Percentage of Q-OPF Scenarios Without μ Correction Out of 1000 Demand Scenarios.

| | 3-bus | 5-bus | 118-bus | 300-bus |
|---|---|---|---|---|
| DC Q-OPF | 89.8 | 77.0 | 90.2 | 71.3 |
| AC Q-OPF | 95.2 | 88.6 | 85.1 | 67.3 |

Removing the $\mu$-modification also led to highly inconsistent algorithmic behavior. In some instances, the objective value appeared to converge to the correct optimum (Fig. 17a), yet the corresponding gradient and complementarity conditions were still violated (Fig. 18), rendering the solution infeasible. In other instances, the solver did not even approach the optimum and instead settled on a suboptimal point (Fig. 17b).

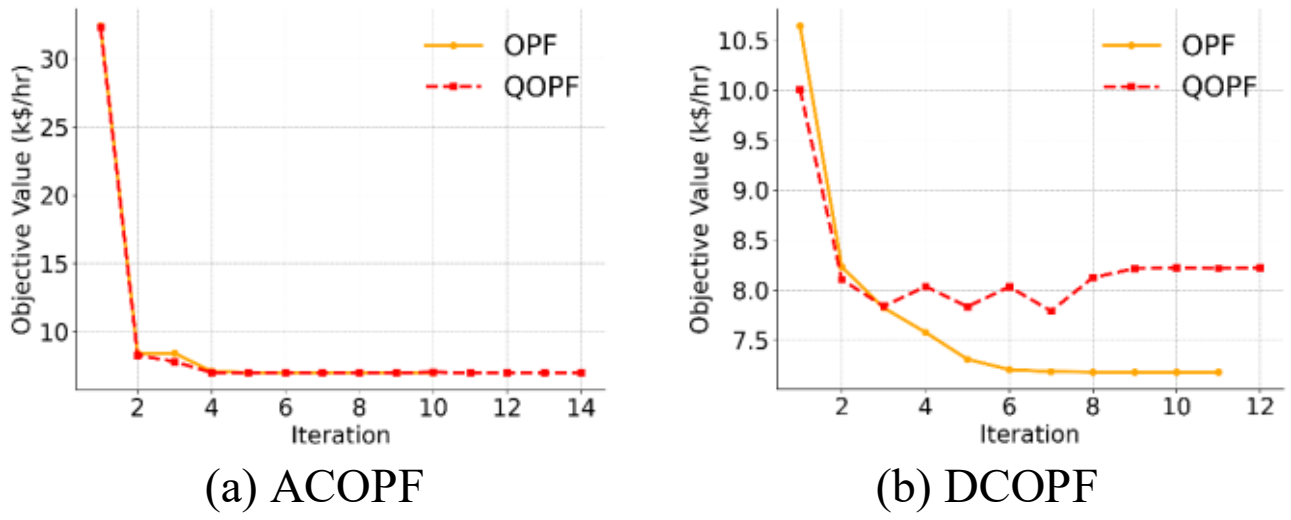

(a) ACOPF (b) DCOPF

Fig. 17: Objective function comparison of OPF and Q-OPF in 3-bus system.

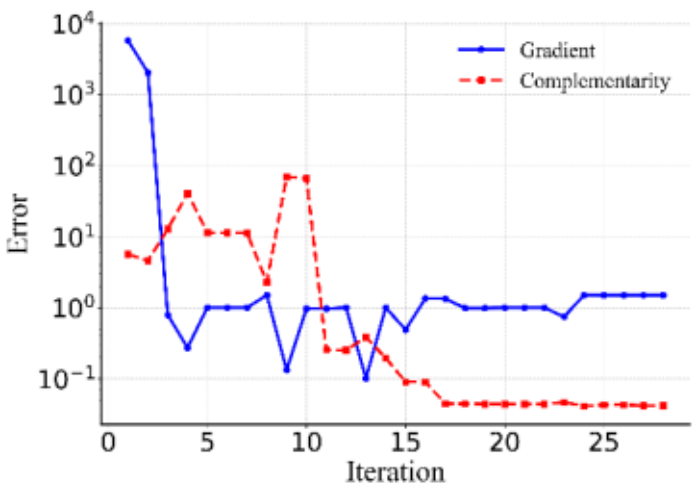

Fig. 18: Gradient and complementarity error of Q-OPF for 3-bus system.

## V. Conclusion

This paper presented a quantum-inspired approach for solving OPF using a variational quantum circuit, with IPM as the core optimization method. The goal was to explore the application of quantum computing in power systems optimization and address its associated challenges. A comparative analysis of HHL, VQLS, and CVQLS showed that CVQLS is the most suitable for OPF, especially when handling ill-conditioned matrices such as the IPM Hessian.

We found shallow ansatz performed well, while deeper ones suffered from overfitting and trainability issues, emphasizing the need for careful parameter control. To improve convergence, we suggested initializing each IPM cycle with the optimized parameters from the previous one. Among various optimizers, CVQLS showed stable performance with Adam, whereas VQLS was unstable across different choices. We also recommended adjusting the central path parameter $\mu$ in IPM to prevent convergence to suboptimal solutions.

Although CVQLS shows promise, its application is currently limited by the capabilities of existing quantum hardware, particularly for larger systems. Our results from testing the proposed QOPF on a real quantum computer using a 2-bus system revealed key limitations. Due to the lack of hybrid circuit support, each quantum circuit had to be submitted separately, resulting in long queue times. Additionally, hardware noise reduced solution accuracy. To address these constraints, we used a quantum simulator for larger cases and tested the approach on IEEE 3-, 5-, 118-, and 300-bus systems, demonstrating satisfactory results.

Future research should focus on enhancing the scalability of CVQLS for larger systems. Exploring advanced ansatz design, enhanced initialization techniques, and adaptive methods for tuning classical parameters could further strengthen the robustness of Q-OPF. These efforts will help optimize the integration of quantum computing into power systems, promoting cost-effective and resilient operations under uncertainty.

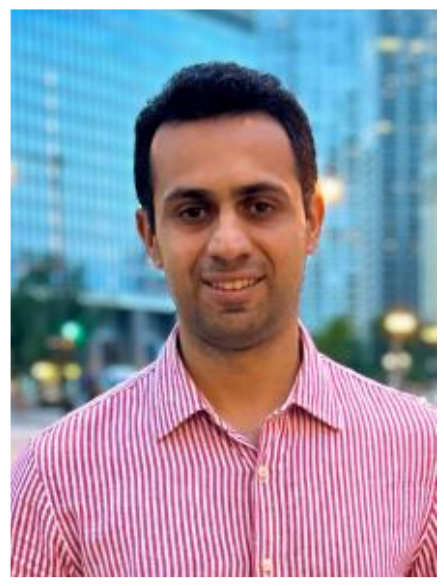

**Farshad Amani** (Graduate Student Member, IEEE) received his B.Sc. degree in Electrical Engineering from the Iran University of Science and Technology, Tehran, Iran, in 2017, and his M.Sc. degree in Power Systems Engineering from the Sharif University of Technology, Tehran, Iran, in 2020. He has been pursuing a Ph.D. in Electrical Engineering at Louisiana State University, Baton Rouge, LA, USA, since 2022. His research interests include optimization, quantum computing, quantum optimization, machine learning, smart grids, power system reliability and resilience, and the operation and planning of power distribution systems

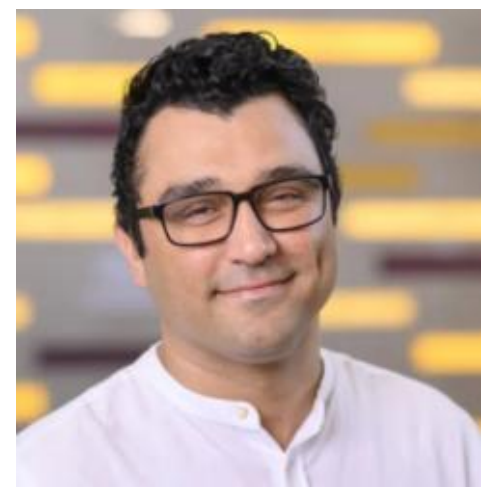

**Amin Kargarian** (Senior Member, IEEE) received the M.Sc. degree from Shiraz University, Iran, 2010 and the Ph.D. in electrical and computer engineering from Mississippi State University, Starkville, MS, USA, in 2014. From 2014 to 2015, he was a Postdoctoral Research Associate with the Department of Electrical and Computer Engineering, Carnegie Mellon University, Pittsburgh, PA, USA. He is currently an Associate Professor with the Electrical and Computer Engineering Department, at Louisiana State University, Baton Rouge, LA, USA. His research interests include optimization, machine learning, quantum computing, and their applications to power systems.